\documentclass {article} 
\usepackage[pctex32]{graphics}
\usepackage {graphicx}
\usepackage{amsmath,amsxtra,amssymb,latexsym, amscd}
\usepackage[mathscr]{eucal}

\makeindex
\begin{document}
\parindent=1.05cm 
\setlength{\baselineskip}{14truept}
\setcounter{page}{1}
\makeatletter 
\title{}
\date{} 
\maketitle
\pagestyle{plain}
\pagestyle{myheadings}
\markboth{\footnotesize The extending for composite Skyrme model }{\footnotesize P. T. Tuyen, D. Q. Tuan }
\begin{center}
\vskip-3cm 
{\large \bf 
THE EXTENDING FOR COMPOSITE SKYRME MODEL
}
\end{center}
\centerline{Pham Thuc Tuyen \footnote{tuyenpt@coltech.vnu.vn (or) phamthuc.tuyen@gmail.com}, Do Quoc Tuan \footnote{do.tocxoan@gmail.com} }
\begin{center}{\it
Department of Physics, Hanoi University of Science, \\ 
334 Nguyen Trai, Thanh Xuan, Hanoi, Vietnam.}
\end{center}
{\small \noindent {\bf Abstract.} \it
In this paper, we have extended the composite Skyrme model proposed by H. Y. Cheung, F. Gursey to some respects of theoretical physics such as supersymmetry, gravitation.
}

\section{Introduction}	
The Skyrme model [2] was proposed to repair weak points of QCD at low
energy (in GeV). Time by time, it really become a phenomenological
model of baryons. In 1990, H. Y. Cheung, F. Gursey [1] gave the general Skyrme model in which the sigma model [3] and the $V-T$ model [9] were two special cases corresponding
to \emph{n=1} and \emph{n=2}. When Cheung and Gursey constructed composite Skyrme model, they also didn't let the mass of pion into the Lagrangian density [1] because of its spontaneous symmetry breaking [3, 4]. However,  datum of prediction are not really axact. However, datum of prediction are not really axact [1]. So, we have proposed the Lagrangian density with the pion's mass. In 1984, E. A. Bergshoeff, R. I. Nepomachie, H. J. Schnitzer proposed Skyrmion of four-dimensional supersymmetric non-linear sigma model [5]. However, there were not papers that concerned with an extending the composite Skyrme model to the supersymmetry. Thus, we have performed to extend the commposite Skyrme model into supersymmetry. Recently, the Einstein-Skyrme system has been studied by several authors [6, 7, 8]. And then the skyrmion to be the gravitating skyrmion. All of these study were related to the non-linear Skyrme model but were not related to the composite Skyrme model. Thus, we have constructed the gravitating composite skyrmion.

\section{The mass of pion}
The Lagrangian density is defined as
\begin{equation}
{\cal L}_n ^{pion}  = {\cal L}_n  + \frac{1}{{8n}}m_\pi ^2 F_\pi ^2 \left[ {Tr\left( {U^n } \right) - 2} \right].
\end{equation}
\vskip0.2cm
Putting the hedgehog solution given by Skyrme [2] $U_o \left( r \right) = \exp \left[ {i\tau .\hat rF\left( r \right)} \right]$ into the Lagrangian density (1), we have
\begin{equation}
{\cal L}_n ^{pion}  = {\cal L}_n  + \frac{1}{{4n}}m_\pi ^2 F_\pi ^2 \left( {\cos nF - 1} \right),
\end{equation}
where 
$\tau $'s are Pauli's matrices, $\hat r = \frac{{\vec r}}{{\left| {\vec r} \right|}}$ and
\begin{equation}
{\cal L}_n  = \frac{{F_\pi ^2 }}{8}\left[ {\left( {\frac{{dF}}{{dr}}} \right)^2  + 2\frac{{\sin ^2 nF}}{{n^2 r^2 }}} \right] + \frac{{\sin ^2 nF}}{{2e^2 n^2 r^2 }}\left[ {2\left( {\frac{{dF}}{{dr}}} \right)^2  + \frac{{\sin ^2 nF}}{{n^2 r^2 }}} \right].
\end{equation}
\vskip0.2cm
The static energy of chiral field is
\begin{equation}
{\cal E}_n ^{pion}  = \frac{{F_\pi  }}{e}A_n  + \frac{1}{{e^3 F_\pi  }}\Theta _n,
\end{equation}
where
\begin{equation}
A_n  = \int\limits_0^\infty  {4\pi \tilde r^2 \left\{ {\left[ {\frac{{F'^2 }}{8} + \frac{{\sin ^2 nF}}{{4n^2 \tilde r^2 }}} \right] + \frac{{e^2 F_\pi ^4 \sin ^2 nF}}{{n^2 \tilde r^2 }}\left[ {F'^2  + \frac{{\sin ^2 nF}}{{2n^2 \tilde r^2 }}} \right]} \right\}} d\tilde r,
\end{equation}
\begin{equation}
\Theta _n  = \int\limits_0^\infty  {\frac{{\pi \tilde r^2 m_\pi ^2 }}{n}} \left( {\cos nF - 1} \right)d\tilde r,
\end{equation}
and $\tilde r = eF_\pi  r$ is a dimensionless variable.
\vskip0.2cm
Now, we shall quantize the hedgehog solution ( or $\it soliton$) by collective coordinates $A\left( t \right) = a_o \left( t \right) + i\vec a\left( t \right)\vec \tau $ with coditions $a_\mu  a_\mu   = 1$ and $a_\mu  \dot a_\mu   = 0$ $\left( {\mu  = 0,1,2,3} \right)$. By this way, solitons being real partilces. Different quantizations make different real particles. With the $A\left( t \right)$ matrix, the $U^n \left( {x,t} \right)$ will transform as $U^n \left( {x,t} \right) = A\left( t \right)U_0^n \left( x \right)A^{ - 1} \left( t \right)$. We put $U^n \left( {x,t} \right)$ into Eq. (1) then the Lagrangian is
\begin{equation}
L_n  =  - {\cal E}_n ^{pion}  + \Gamma _n Tr\left[ {\partial _0 A\left( t \right)\partial _0 A^{ - 1} \left( t \right)} \right] =  - {\cal E}_n ^{pion}  + 2\Gamma _n \sum\limits_{\mu  = 0}^3 {\dot a_\mu ^2 },
\end{equation}
where
\begin{equation}
\Gamma _n  = \frac{1}{{e^3 F_\pi  }}\Phi _n,
\end{equation}
\begin{equation}
\Phi _n  = \frac{{2\pi }}{{3n^2 }}\int\limits_0^\infty  {\tilde r^2 \sin ^2 nF\left\{ {1 + 4\left[ {\left( {\frac{{dF}}{{d\tilde r}}} \right)^2  + \frac{{\sin ^2 nF}}{{n^2 \tilde r^2 }}} \right]} \right\}} d\tilde r.
\end{equation}
\vskip0.2cm
The conjugate momenta are defined as
\begin{equation}
\pi _\mu   = \frac{{\partial L_n }}{{\partial \dot a_\mu  }} = \partial \left[ { - {\cal E}_n ^{pion}  + 2\Gamma _n \sum\limits_{\mu  = 0}^3 {\dot a_\mu ^2 } } \right]/\partial \dot a_\mu   = 4\Gamma _n \dot a_\mu.
\end{equation}
\vskip0.2cm
From Eq. (7), we can get the Hamiltonian  
\begin{equation}
H_n  = \pi _\mu  \dot a_\mu   - L_n  = 4\Gamma _n \dot a_\mu  \dot a_\mu   - L_n  = {\cal E}_n ^{pion}  + 2\Gamma _n \dot a_\mu  \dot a_\mu   = {\cal E}_n ^{pion}  + \left( {\sum\limits_{\mu  = 0}^3 {\pi _\mu ^2 } } \right)/8\Gamma _n.
\end{equation}
\vskip0.2cm
Canonical quantizing momenta $\pi _\mu   =  - i\partial /\partial a_\mu  $ takes the Hamiltonian as
\begin{equation}
H_n  = {\cal E}_n ^{pion}  + \left( {1/8\Gamma _n } \right)\sum\limits_{\mu  = 0}^3 {\left( { - \partial ^2 /\partial a_\mu ^2 } \right)}.
\end{equation}
\vskip0.2cm
The isotopic vector $\vec c$ is defined as 
\begin{equation}
\dot A^\dag  A = \left( {\dot a_0  - i\dot a_i \tau _i } \right)\left( {a_0  + ia_i \tau _i } \right) = \dot a_0 a_0  + \dot \vec a\vec a - i\left( {a_0 \dot \vec a - \dot a_0 \vec a + \vec a \times \dot \vec a} \right)\vec \tau  =  - i\vec c\vec \tau.
\end{equation}
\vskip0.2cm
The square isotopic vector $\vec c$ is
\begin{equation}
\vec c^2  = \vec a^2 \dot a_0^2   + \dot \vec a^2 a_0^2  + \vec a^2 \dot \vec a^2  - (\dot \vec a\vec a)^2  - 2\dot a_0 a_0 \dot \vec a\vec a = \vec a^2 \dot a_0^2  + \dot \vec a^2 a_0^2  + \vec a^2 \dot \vec a^2  - (\dot \vec a\vec a)^2  + 2(\dot \vec a\vec a)^2  = \dot a_\mu  \dot a_\mu.
\end{equation}
\vskip0.2cm
In the configuration of hedgehog, the spin and isospin rotation are equivalent. We can homogenize $\vec c^2$ with the square of spin and isospin
\begin{equation}
\vec c^2  = \frac{1}{{16\Gamma _n^2 }}\vec J^2  = \frac{1}{{16\Gamma _n^2 }}\vec T^2.
\end{equation}
\vskip0.2cm
Thus, we have the Hamiltonian
\begin{equation}
H_n  = {\cal E}_n ^{pion}  + \frac{1}{{8\Gamma _n }}\vec J^2  = {\cal E}_n ^{pion}  + \frac{1}{{8\Gamma _n }}\vec T^2,
\end{equation}
where $\vec J$ is a spin and $\vec T$ is an isospin. Eigenvalues of Hamiltonian (16) are
\begin{equation}
M_n  = E_n  = \frac{{F_\pi  }}{e}A_n  + \frac{1}{{e^3 F_\pi  }}\Theta _n  + \frac{{l\left( {l + 2} \right)}}{{8\Gamma _n }},
\end{equation}
where $l=2J=2I$ and $(J, I)$ are respectively the isospin and spin quantum numbers. So, the mass of nucleon $\left( {J = 1/2} \right)$ and delta $\left( {J = 3/2} \right)$  are given by
\begin{equation}
M_N  = \frac{{F_\pi  }}{e}A_n  + \frac{1}{{e^3 F_\pi  }}\Theta _n  + \frac{{3e^3 F_\pi  }}{{8\Phi _n }},
\end{equation}
\begin{equation}
M_\Delta   = \frac{{F_\pi  }}{e}A_n  + \frac{1}{{e^3 F_\pi  }}\Theta _n  + \frac{{15e^3 F_\pi  }}{{8\Phi _n }}.
\end{equation}
\vskip0.2cm
From Eqs. (18) and (19), we can get
\begin{equation}
F_\pi   = \frac{{2\left( {M_\Delta - M_N } \right)\Phi _n }}{{3e^3 }},
\end{equation}
\begin{equation}
e = \sqrt[4]{{\frac{{4\left( {M_\Delta - M_N } \right)^3 \Phi _n^2 A_n }}{{6M_N \left( {M_\Delta - M_N } \right)^2 \Phi _n - 9\Theta _n - 3\left( {M_\Delta - M_N } \right)^2 \Phi _n }}}}.
\end{equation}
\vskip0.2cm
We obtain the nonlinear differential equation of $F\left( \tilde r \right)$ from the minimum condition of the chiral field's energy: $\delta _F {\cal E}_n ^{pion} = 0$
\begin{equation}
\left( {\frac{{\tilde r^2 }}{4} + \frac{2}{{n^2 }}\sin ^2 nF} \right)F'' + \frac{{\tilde r}}{2}F' + \frac{{\sin 2nF}}{n}\left( {F'^2  - \frac{{\sin ^2 nF}}{{n^2 \tilde r^2 }} - \frac{1}{4}} \right) - \beta \tilde r^2 \sin nF = 0,
\end{equation}
where
\begin{equation}
\beta  = \frac{{m_\pi ^2 }}{{4e^2 F_\pi ^2 }}.
\end{equation}
\vskip0.2cm
We see that Eq. (23) contain the constant of $e$ and $F_\pi $ but their values are not known. They can be found after we solve Eq. (22). Thus, we must use the special way to solve Eq. (22). Firstly, we assume the value of $\beta^* $, put it into Eq. (22) to solve this equation. When numerical solution of Eq. (22) is found, values of $F_\pi$, $e$ and $\beta^{**} $ will be found from Eqs. (20), (21) and (23). After, we shall compare two $\beta^* $ and $\beta^{* *}$ until they equal approximately together. For the case of $n=4$, we need the value of $\beta $ is 0.0025.
\vskip0.2cm
After constructing coposite Skyrme model with pion's mass, we shall apply this model to calculate some static properties of nucleon (proton and neutron) [1] in a case of $n=4$.  And numerical results of static properties of nucleon are described in Fig. (2) and the below table 
\vskip0.2cm
\centerline{\bf\tiny 
NUMERICAL RESULTS TABLE IN A CASE OF N=4
}
\vskip0.4cm
\hspace{2cm} {\bf Quantity} \hspace{1.7cm} {\bf Prediction} \hspace{1.2cm} {\bf Experiment}
\vskip0.2cm
\hspace{2cm} $e$ \hspace{3.8cm} 2.87 \hspace{3cm} -
\vskip0.2cm
\hspace{2cm} $F_\pi $(MeV) \hspace{2.2cm} 446.56 \hspace{2.5cm} 186
\vskip0.2cm
\hspace{2cm} $\left\langle {r^2 } \right\rangle _{E,I = 0}^{1/2} $(fm) \hspace{1.6cm} 0.21 \hspace{2.8cm} 0.71
\vskip0.2cm
\hspace{2cm} $\left\langle {r^2 } \right\rangle _{E,I = 1}^{1/2} $(fm) \hspace{1.5cm} 2.26 \hspace{2.9cm} 0.88
\vskip0.2cm
\hspace{2cm} $\left\langle {r^2 } \right\rangle _{M,I = 0}^{1/2} $(fm) \hspace{1.6cm} 1.0 \hspace{2.9cm} 0.81
\vskip0.2cm
\hspace{2cm} $\left\langle {r^2 } \right\rangle _{M,I = 1}^{1/2} $(fm) \hspace{1.5cm} 2.26 \hspace{2.8cm} 0.8
\vskip0.2cm
\hspace{2cm} $\mu _{pro} $(mag) \hspace{2.2cm} 1.64 \hspace{2.8cm} 2.79
\vskip0.2cm
\hspace{2cm} $\mu _{neu} $(mag) \hspace{2cm} -1.57 \hspace{2.6cm} -1.91
\vskip0.2cm
\hspace{2cm} $\left| {\frac{{\mu _{pro} }}{{\mu _{neu} }}} \right|$ \hspace{2.9cm} 1.04 \hspace{2.8cm} 1.46
\vskip0.2cm
\hspace{2cm} $g_A $ \hspace{3.45cm} 1.22 \hspace{2.8cm} 1.23

\section{The supersymmetric composite Skyrme model}

 The Lagrangian density of composite Skyrme model is defined as
\begin{equation}
{\cal L}_n  =  - \frac{{f_\pi ^2 }}{{16n^2 }}Tr\left( {\partial
_\mu  U^{ - n} \partial ^\mu  U^n } \right) + \frac{1}{{32e^2 n^4
}}Tr\left( {\left[ {U^{ - n} \partial _\mu  U^n ,U^{ - n} \partial
_\nu  U^n } \right]^2 } \right) ,
\end{equation}
where $f_\pi  $ is the pi-meson decay constant, \emph{\textbf{e}} is a
dimensionless parameter.
\vskip0.2cm 
The ordinary derivatives in the Lagrangian density (24) is replaced into the covariant derivatives
\begin{equation}
D_\mu  U^n  = \partial _\mu  U^n  - iV_\mu  U^n \tau _3  ,
\end{equation}
then  Eq. (24) is
\begin{equation}
{\cal L}_n  =  - \frac{{f_\pi ^2 }}{{16n^2 }}Tr\left( {D_\mu  U^{ - n} D^\mu  U^n } \right) + \frac{1}{{32e^2 n^4 }}Tr\left( {\left[ {U^{ - n} D_\mu  U^n ,U^{ - n} D_\nu  U^n } \right]^2 } \right) .
\end{equation}
\vskip0.2cm
Eq. (26) is invariant under the local  $U\left( 1 \right)_R $ and the global $SU\left( 2 \right)_L $  transformations
\begin{equation}
U^n \left( r \right) \to AU^n \left( x \right)e^{i\lambda \left( r \right)\tau _3 },  A \in SU\left( 2 \right)_L ,
\end{equation}
\begin{equation}
V_\mu  \left( r \right) \to V_\mu  \left( r \right) + \partial _\mu  \lambda \left( r \right) .
\end{equation}
\vskip0.2cm 
The gauge field $V_\mu  \left( r \right)$  in Eq. (28) is defined as
\begin{equation}
V_\mu   =  - \frac{i}{{2n}}Tr\left( {U^{ - n} \partial _\mu  U^n \tau _3 } \right) .
\end{equation}
\vskip0.2cm
We parametrize the $SU\left( 2 \right)$ matrix $U^n \left( r \right)$ in terms of the complex scalars $A_i $  
\begin{equation}
U^n \left( r \right) = \left( {\begin{array}{*{20}c}
   {A_1 } & { - A_2^* }  \\
   {A_2 } & {A_1^* }  \\
\end{array}} \right) .
\end{equation}
\vskip0.2cm
From the unitary constraint, we have 
\begin{equation}
U^\dag  U = 1 \Rightarrow \bar A^i A_i  = A_1^* A_1  + A_2^* A_2  = 1 .
\end{equation}
\vskip0.2cm
Eq. (25) can be rewritten as 
\begin{equation}
D_\mu  \left( {\begin{array}{*{20}c}
   {A_1 } & { - A_2^* }  \\
   {A_2 } & {A_1^* }  \\
\end{array}} \right) = \partial _\mu  \left( {\begin{array}{*{20}c}
   {A_1 } & { - A_2^* }  \\
   {A_2 } & {A_1^* }  \\
\end{array}} \right) - iV_\mu  \left( {\begin{array}{*{20}c}
   {A_1 } & {A_2^* }  \\
   {A_2 } & { - A_1^* }  \\
\end{array}} \right)
\end{equation}
or
\begin{equation}
D_\mu  A_i  = \left( {\partial _\mu   - iV_\mu  } \right)A_i ,
\end{equation}
\begin{equation}
D_\mu  \bar A_i  = \left( {\partial _\mu   + iV_\mu  } \right)\bar A_i .
\end{equation}
\vskip0.2cm
Then, the form of gauge field (29) is
\begin{equation}
V_\mu  \left( r \right) =  - \frac{i}{{2n}}\bar A^i \mathord{\buildrel{\lower3pt\hbox{$\scriptscriptstyle\leftrightarrow$}} 
\over \partial } A_i .
\end{equation}
\vskip0.2cm
With the matrix $U^n \left( r \right)$ defined in Eq. (30) and the gauge field defined in Eq. (35), the supersymmetric Lagrangian density is
\begin{equation}
{\cal L}_n  =  - \frac{{f_\pi ^2 }}{{8n^2 }}\bar D_\mu  \bar AD^\mu  A - \frac{1}{{16e^2 n^2 }}F_{\mu \nu }^2 \left( V \right) ,
\end{equation}
where
\begin{equation}
F_{\mu \nu } \left( V \right) = \partial _\mu  V_\nu   - \partial _\nu  V_\mu .
\end{equation}
\vskip0.2cm
To supersymmetrise composite Skyrme model, we extend $A_i $ to chiral scalar multiplets $\left( {A_i ,\psi _{\alpha i} ,F_i } \right)$ $\left( {i,\alpha  = 1,2} \right)$ and the vector $V_\mu  \left( r \right)$ to real vector multiplets $\left( {V_\mu  ,\lambda _\alpha  ,D} \right)$. Here, the fields $F_i $ are complex scalars, $D$ is real scalar, $\psi _{\alpha i} $, $\lambda _\alpha  $ are Majorana two-component spinors. $\psi _{\alpha i} $ corresponds to a left-handed chiral spinor, $\bar \psi ^{\alpha i}  = \left( {\psi _i^\alpha  } \right)^* $  corresponds to a right-handed one. The supersymmetric Lagrangian density is given by
\begin{center}
${\cal L}_{susy}  = \frac{{f_\pi ^2 }}{{8n^2 }}\left[ { - D^\mu  \bar A^i D_\mu  A_i  - \frac{1}{2}i\bar \psi ^{\dot \alpha i} \left( {\sigma _\mu  } \right)_{\alpha \dot \alpha } \mathord{\buildrel{\lower3pt\hbox{$\scriptscriptstyle\leftrightarrow$}} 
\over D} ^\mu  \psi _i^\alpha   + \bar F^i F_i  - }\right.$
\end{center}
\begin{center}
$\left. { - i\bar A^i \lambda ^\alpha  \psi _{\alpha i}  + iA_i \bar \lambda ^{\dot \alpha } \bar \psi _{\dot \alpha }^i  + D\left( {\bar A^i A_i  - 1} \right)} \right] +$ 
\end{center}
\begin{equation}
+ \frac{1}{{8e^2 n^2 }}\left[ { - \frac{1}{2}F_{\mu \nu }^2  - i\bar \lambda ^{\dot \alpha } \left( {\sigma ^\mu  } \right)_{\dot \alpha }^\alpha  \partial _\mu  \lambda _\alpha   + D^2 } \right].
\end{equation}
\vskip0.2cm
It is invariant under the following set of supersymmetric transformations [5]
\begin{equation}
\delta A_i  =  - \varepsilon ^\alpha  \psi _{\alpha i} ,
\end{equation}
\begin{equation}
\delta \psi _{\alpha i}  =  - i\bar \varepsilon ^{\dot \alpha } \left( {\sigma ^\mu  } \right)_{\alpha \dot \alpha } D_\mu  A_i  + \varepsilon _\alpha  F_i ,
\end{equation}
\begin{equation}
\delta F_i  =  - i\bar \varepsilon ^{\dot \alpha } \left( {\sigma ^\mu  } \right)_{\dot \alpha }^\alpha  D_\mu  \psi _{\alpha i}  - i\bar \varepsilon ^{\dot \alpha } A_i \bar \lambda _{\dot \alpha } ,
\end{equation}
\begin{equation}
\delta V_\mu   =  - \frac{1}{2}i\left( {\sigma _\mu  } \right)^{\alpha \dot \alpha } \left( {\bar \varepsilon _{\dot \alpha } \lambda _\alpha   + \varepsilon _\alpha  \bar \lambda _{\dot \alpha } } \right) ,
\end{equation}
\begin{equation}
\delta \lambda _\alpha   = \varepsilon ^\beta  \left( {\sigma ^{\mu \nu } } \right)_{\beta \alpha } F_{\mu \nu }  + i\varepsilon _\alpha  D ,
\end{equation}
\begin{equation}
\delta D = \frac{1}{2}\left( {\sigma ^\mu  } \right)_{\alpha \dot \alpha } \partial _\mu  \left( {\bar \varepsilon ^{\dot \alpha } \lambda ^\alpha   - \varepsilon ^\alpha  \bar \lambda ^{\dot \alpha } } \right) .
\end{equation}
\vskip0.2cm
The field equations and their supersymmetric transformations lead to the following constraints [5]
\begin{equation}
\bar A^i A_i  = 0 ,
\end{equation}
\begin{equation}
\bar A^i \psi _{\alpha i}  = 0 ,
\end{equation}
\begin{equation}
\bar A^i F_i  = 0
\end{equation}
and following algebraic expressions [5]
\begin{equation}
V_\mu   =  - \frac{1}{2}\left\{ {i\bar A^i \mathord{\buildrel{\lower3pt\hbox{$\scriptscriptstyle\leftrightarrow$}} 
\over \partial } _\mu  A_i  + \left( {\sigma _\mu  } \right)^{\alpha \dot \alpha } \bar \psi _{\dot \alpha }^i \psi _{\alpha i} } \right\} ,
\end{equation}
\begin{equation}
\lambda _\alpha   =  - i\left\{ {\bar F^i \psi _{\alpha i}  + i\left( {\sigma ^\mu  } \right)_{\alpha \dot \alpha } \left( {D_\mu  A_i } \right)\bar \psi ^{\dot \alpha i} } \right\} ,
\end{equation}
\begin{equation}
D = D^\mu  \bar A^i D_\mu  A_i  + \frac{1}{2}i\bar \psi ^{\dot \alpha i} \left( {\sigma ^\mu  } \right)_{\alpha \dot \alpha } \left( {\mathord{\buildrel{\lower3pt\hbox{$\scriptscriptstyle\leftrightarrow$}} 
\over D} _\mu  \psi _i^\alpha  } \right) - \bar F^i F_i .
\end{equation}
\vskip0.2cm
Setting $\psi _\alpha   = F_i  = 0$, Eq. (39) is
\begin{equation}
{\cal L}_{susy}  =  - \frac{{f_\pi ^2 }}{{8n^2 }}\bar D^\mu  \bar AD_\mu  A + \frac{1}{{8e^2 n^2 }}\left[ { - \frac{1}{2}F_{\mu \nu }^2  + \left( {\bar D^\mu  \bar AD_\mu  A} \right)^2 } \right] .
\end{equation}
\vskip0.2cm
We see that in Eq. (51) the second term is fourth-order in time derivatives. However, there are other possible fourth-order terms [5]
\begin{equation}
\square \bar A\square A - \left( {\bar D^\mu  \bar AD_\mu  A} \right)^2 .
\end{equation}
\vskip0.2cm
 Thus, we can add Eq. (52) into Lagrangian density (51)
\begin{center}
${\cal L}_{susy}  =  - \frac{{f_\pi ^2 }}{{8n^2 }}\bar D^\mu  \bar AD_\mu  A + \frac{1}{{8e^2 n^2 }}\left[ {\alpha \left\{ { - \frac{1}{2}F_{\mu \nu }^2  + \left( {\bar D^\mu  \bar AD_\mu  A} \right)^2 } \right\} + } \right.$
\end{center}
\begin{equation}
\left. { + \beta \left\{ {\square \bar A\square A - \left( {\bar D^\mu  \bar AD_\mu  A} \right)^2 } \right\}} \right] .
\end{equation}
\vskip0.2cm
The hedgehog ansatz is defined as [1, 2]
\begin{equation}
U^n \left( r \right) = \cos nf\left( r \right) + i\vec \tau \frac{{\vec r}}
{r}\sin nf\left( r \right) .
\end{equation}
\vskip0.2cm
From Eq. (30), we have
\begin{equation}
A_1  = \cos nf\left( r \right) + i\cos \theta \sin nf\left( r \right) ,
\end{equation}
\begin{equation}
A_2  = ie^{i\varphi } \sin \theta \sin nf\left( r \right) .
\end{equation}
\vskip0.2cm
Inserting Eqs. (55) and (56) into the supersymmetric Lagrangian density (53), we obtain the static energy
\begin{center}
$E = 4\pi \frac{{f_\pi  }}{e}\int {dxx^2 } \left\{ {\frac{1}{{12}}\left( {f'^2  + \frac{{2\sin ^2 nf}}{{n^2 x^2 }}} \right) + \left( {\frac{{\alpha  + \beta }}{{15}}} \right)\left( {f'^2  - \frac{{\sin ^2 nf}}{{n^2 x^2 }}} \right) + } \right.$
\end{center}
\begin{equation}
\left. { + \frac{\beta }{{12}}\left( {f'' + \frac{{2f'}}{{nx}} - \frac{{\sin 2nf}}{{n^2 x^2 }}} \right)} \right\} .
\end{equation}
\vskip0.2cm
From a minimum condition of the hedgehog's energy $\delta _f E = 0$, we obtain the following nonlinear differential equation of $f(x)$
\begin{center}
$- x^2 f'' - 2xf' + \frac{{\sin 2nf}}{n} + \frac{{4\left( {\alpha  + \beta } \right)}}{5}\left[ {\frac{{2f''\sin ^2 nf}}{{n^2 }} - 6x^2 f'^2 f'' - 4xf'^3  + } \right.$
\end{center}
\begin{center}
$\left. { + \frac{{f'^2 \sin 2nf}}{n} + \frac{{\sin ^2 nf\sin 2nf}}{{n^3 x^2 }}} \right] + \beta \left[ {x^2 f^{\left( 4 \right)}  + \frac{{4xf^{\left( 3 \right)} }}{n} - \frac{{4f''\cos 2nf}}{n} + } \right.$
\end{center}
\begin{equation}
\left. { + \frac{{4f'^2 \sin 2nf}}{{n^2 }} - \frac{{4\sin ^2 nf\sin 2nf}}{{n^3 x^2 }}} \right] = 0 ,
\end{equation}
where $x= ef_\pi  r$ is the dimensionless variable.
\vskip0.2cm
Two boundary conditions of  Eq. (58) are defined as
\begin{equation}
f\left( 0 \right) = \pi,
\end{equation}
\begin{equation}
 f\left( \infty  \right) = 0 .
\end{equation}
\vskip0.2cm
Eq. (58) can not give the analytic solutions. So we use numerical method to solve it. 
\section{The gravitational composite Skyrme model}
The Lagrangian composite Skyrme model coupled with gravity can be defined as

\begin{equation}
{\cal L} = {\cal L}_{\cal G}  + {\cal L}_{{\cal C}{\cal S}} 
\end{equation}
where
${\cal L}_{\cal G}  = \frac{1}{{16\pi G}}R$, R is a curvature scalar and G is a Newton constant, 
\begin{equation}
{\cal L}_{{\cal C}{\cal S}}  =  - \frac{{F_\pi ^2 }}{{16n^2 }}g^{\mu \nu } Tr\left( {\partial _\mu  U^n \partial _\nu  U^n } \right) + 
\end{equation}
\begin{equation}
 + \frac{1}{{32e^2 n^4 }}g^{\mu \nu } g^{\rho \sigma } Tr\left[ {\left( {\partial _\mu  U^n } \right)U^{ - n} ,\left( {\partial _\rho  U^n } \right)U^{ - n} } \right]\left[ {\left( {\partial _\nu  U^n } \right)U^{ - n} ,\left( {\partial _\sigma  U^n } \right)U^{ - n} } \right],
\end{equation}
U(x) is an SU(2) chiral field,  $F_\pi  $ is the pion decay constant, e is a dimesionless parameter.
We consider the static spherically symmetric metric given by
\begin{equation}
ds^2  =  - N^2 \left( r \right)C\left( r \right)dt^2  + \frac{1}{{C\left( r \right)}}dr^2  + r^2 \left( {d\theta ^2  + \sin ^2 \theta d\varphi ^2 } \right),
\end{equation}
where
\begin{equation}
C\left( r \right) = 1 - \frac{{2m\left( r \right)}}{r}.
\end{equation}
From the equation
\begin{equation}
ds^2  = g_{\mu \nu } dx^\mu  dx^\nu ,
\end{equation}
we can define tensor metric  $g_{\mu \nu } $
\begin{equation}
g_{\mu \nu }  = \left( { - N^2 \left( r \right)C\left( r \right),\frac{1}{{C\left( r \right)}},1,1} \right)_{diag} 
\end{equation}
\begin{equation}
\to \sqrt { - g}  = N\left( r \right).
\end{equation}
Inserting the hedgehog ansatz proposed by Skyrme [1] $U\left( r \right) = c{\rm{osf}}\left( {\rm{r}} \right) + i\vec n\vec \tau {\mathop{\rm s}\nolimits} {\rm{inf}}\left( {\rm{r}} \right)$  into the Lagrangian density (1), we can obtain the static energy density for the chiral field
\begin{equation}
{\cal E}_S  = \frac{{F_\pi ^2 }}{8}\left( {Cf'^2  + 2\frac{{\sin ^2 nf}}{{n^2 r^2 }}} \right) + \frac{1}{{2e^2 }}\frac{{\sin ^2 nf}}{{n^2 r^2 }}\left( {2Cf'^2  + \frac{{\sin ^2 nf}}{{n^2 r^2 }}} \right),
\end{equation}
where $\vec \tau $  are Pauli matrices and $\vec n = \frac{{\vec r}}{r}$ .\\
Introducing dimensionless variables, they are useful in solving a field equation
\begin{equation}
x = eF_\pi  r ,
\end{equation}
\begin{equation}
\mu \left( x \right) = eF_\pi  m\left( r \right).
\end{equation}
In terms of $x$  and $\mu \left( x \right)$, the static energy can be written as
\begin{equation}
E_S  = \int {\sqrt { - g} } {\cal E}_S 4\pi r^2 dr ,
\end{equation}
or
\begin{equation}
E_S  = 4\pi \frac{{F_\pi  }}{e}\Gamma,
\end{equation}
where
\begin{equation}
\Gamma  = \int\limits_0^\infty  {\left\{ {\frac{1}{8}\left( {ef'^2  + 2\frac{{\sin ^2 nf}}{{n^2 x^2 }}} \right) + \frac{{\sin ^2 nf}}{{2n^2 x^2 }}\left( {2Cf'^2  + \frac{{\sin ^2 nf}}{{n^2 x^2 }}} \right)} \right\}Nx^2 dx}.
\end{equation}
The energy tensor is defined as 
\begin{equation}
T_{ij}  = \frac{2}{{\sqrt { - g} }}\left[ {\frac{{\partial \left( {\sqrt { - g} {\cal L}_{\cal G} } \right)}}{{\partial g^{ij} }} - \frac{\partial }{{\partial x^k }}\frac{{\partial \left( {\sqrt { - g} {\cal L}_{\cal G} } \right)}}{{\partial g_k^{ij} }}} \right],
\end{equation}
where
\begin{equation}
\frac{{\partial \sqrt { - g} }}{{\partial g^{ij} }} =  - \frac{1}{2}\sqrt { - g} g_{ij}.
\end{equation}
Inserting the form of Lagrangian density (61) into Eq. (76), we can get
\begin{equation}
T_{ij}  =  - \frac{1}{{2n^2 }}F_\pi ^2 Tr\left( {L_i L_j  - \frac{1}{2}g_{ij} L_\mu  L^\mu  } \right) + \frac{1}{{8n^2 }}Tr\left( {F_{i\mu } F_{j\nu } g^{\mu \nu }  - \frac{1}{4}g_{ij} F_{\mu \nu } F^{\mu \nu } } \right),
\end{equation}
where
\begin{equation}
L_\mu   = U^{ - n} \partial _\mu  U^n ,
\end{equation}
\begin{equation}
F_{\mu \nu }  = \left[ {L_\mu  ,L_\nu  } \right].
\end{equation}
The Einstein's gravitational field equation is
\begin{equation}
R_{ij}  - \frac{1}{2}g_{ij} R = 8\pi GT_{ij} ,
\end{equation}
or
\begin{equation}
R_{ij} g^{ij}  - \frac{1}{2}g_{ij} Rg^{ij}  = 8\pi GT_{ij} g^{ij} .
\end{equation}
For i = j = 0,1 we can get two equations for two unknown variables  $N(x)$, $\mu(x) $
\begin{equation}
R_{00} g^{00}  - \frac{1}{2}g_{00} Rg^{00}  = 8\pi GT_{00} g^{00}
\end{equation}
\begin{equation}
\to N' = \frac{\alpha }{4}\left( {x + \frac{{8\sin ^2 nf}}{{n^2 x}}} \right)Nf'^2 ,
\end{equation}
\begin{equation}
R_{11} g^{11}  - \frac{1}{2}g_{11} Rg^{11}  = 8\pi GT_{11} g^{11}
\end{equation}
\begin{equation}
\to \mu ' = \frac{\alpha }{8}\left[ {\left( {x^2  + 8\frac{{\sin ^2 nf}}{{n^2 }}} \right)Cf'^2  + 2\frac{{\sin ^2 nf}}{{n^2 }} + 4\frac{{\sin ^4 nf}}{{x^2 n^4 }}} \right].
\end{equation}
where  $\alpha  = 4\pi GF_\pi ^2 $  is the coupling constant.
The variation of the static energy of chiral field with respect to the profile $f(x)$ leads to the field equation
\begin{equation}
\delta _f {\cal E}_S  = 0 ,
\end{equation}
\begin{center}
$NC\left( {x^2  + 8\frac{{\sin ^2 nf}}{{n^2 }}} \right)f'' + \left( {x^2  + 8\frac{{\sin ^2 nf}}{{n^2 }}} \right)N'Cf' - \left( {1 + 4\frac{{\sin ^2 nf}}{{n^2 x^2 }} + 4Cf'^2 } \right)N\sin 2nf + $
\end{center}
\begin{equation}
+ 2\left( {x + 4\sin 2nff'} \right)NCf'^2  + 2\left( {1 + 8\frac{{\sin ^2 nf}}{{n^2 x^2 }}} \right)\left( {\mu  - \mu 'x} \right)Nf' = 0.
\end{equation}
Boundary conditions of this equation are [1, 3]
\begin{equation}
f\left( 0 \right) = \pi ,
\end{equation}
\begin{equation}
f\left( \infty  \right) = 0.
\end{equation}
To describe nucleon states, we need to quantizate classical skyrmion. The useful method that is used is a collective quantization. Consider collective coordinates are $A\left( t \right) = a_0 \left( t \right) + i\vec \tau \vec a\left( t \right)$  with conditions $a_\mu  a_\mu   = 1$  and $a_\mu  \dot a_\mu   = 0$ ($\mu  = 0, 1, 2, 3$ ). By this way, skyrmions to be real particles (as nucleon). With A(t), the  $U^n \left( {x,t} \right)$ will trasform as
\begin{equation}
U^n \left( {x,t} \right) = A\left( t \right)U_0^n \left( x \right)A^{ - 1} \left( t \right).
\end{equation}
Putting Eq. (90) into Lagrangian of Einstein-Skyrme system (61), we can get
\begin{equation}
{\cal L} =  - M + \lambda Tr\left( {\dot A\dot A^{ - 1} } \right) =  - M + 2\lambda \sum\limits_{\mu  = 0}^3 {\dot a_\mu ^2 } ,
\end{equation}
where $M=E_S$ is classical skyrmion's mass (in the unit $c = \hbar  = 1$)
and
\begin{equation}
\lambda  = \frac{{2\pi }}{{3F_\pi  e^3 }}\Lambda ,
\end{equation}
\begin{equation}
\Lambda  = \int\limits_0^\infty  {\frac{1}{{NC}}} \left[ {1 + 4\left( {Cf' + \frac{{\sin ^2 nf}}{{n^2 x^2 }}} \right)\frac{{x^2 \sin ^2 nf}}{{n^2 }}} \right]dx .
\end{equation}
Following [3], the Hamiltonian is defined as
\begin{equation}
H = M + \frac{1}{{8\lambda }}\sum\limits_{\mu  = 0}^3 {\left( { - \frac{{\partial ^2 }}{{\partial a_\mu ^2 }}} \right)}.
\end{equation}
Eigenvalues of the Hamiltonian (94) are
\begin{equation}
E = M + \frac{{l\left( {l + 2} \right)}}{{8\lambda }},
\end{equation}
where $l=2I=2J$ and ($I, J$) are respectively an isospin and a spin quantum number. For the nucleon ($J=1/2$) and the delta ($J=3/2$) we get
\begin{equation}
M_N  = M + \frac{3}{{8\lambda }},
\end{equation}
\begin{equation}
M_\Delta   = M + \frac{{15}}{{8\lambda }}.
\end{equation}
From Eqs. (73), (74), (92), (93), (96), (97), e and $F_\pi  $  can be found as
\begin{equation}
F_\pi   = \frac{1}{{4\sqrt 6 }}\sqrt[4]{{\frac{{\Lambda \left( {M_\Delta   - M_N } \right)\left( {5M_N  - M_\Delta  } \right)}}{{\pi ^2 \Gamma ^3 }}}},
\end{equation}
\begin{equation}
e = \frac{{16\pi F_\pi  \Gamma }}{{5M_N  - M_\Delta  }}.
\end{equation}
The baryon current is defined as
\begin{equation}
B^\mu   = \frac{{\varepsilon ^{\mu \nu \rho \sigma } }}{{24\pi ^2 n}}\frac{1}{{\sqrt { - g} }}Tr\left( {U^{ - n} \partial _\nu  U^n U^{ - n} \partial _\rho  U^n U^{ - n} \partial _\sigma  U^n } \right).
\end{equation}
From Eq. (90), the baryon current's zeroth component (baryon number density) is found 
\begin{equation}
B^0  =  - \frac{1}{{2\pi ^2 }}\frac{1}{N}\frac{{f'\sin ^2 nf}}{{r^2 }}.
\end{equation}
Using Eqs. (68), (88), (89) the baryon number becomes
\begin{equation}
B = \int\limits_0^\infty  {4\pi r^2 \sqrt { - g} } B^0 dr =  - \frac{2}{\pi }\int\limits_\pi ^0 {\sin ^2 } nfdf = 1.
\end{equation}
The isoscalar mean square radius is defined in terms of the baryon number density
\begin{equation}
\left\langle {r^2 } \right\rangle _{I = 0}  = \int\limits_0^\infty  {r^2 \sqrt { - g} } B^0 4\pi r^2 dr = \int\limits_0^\infty  {4\pi r^4 N} B^0 dr =  - \frac{1}{{\left( {eF_\pi  } \right)^2 }}\frac{2}{\pi }\int\limits_0^\infty  {x^2 f'\sin ^2 nfdx}.
\end{equation}
The isoscalar magnetic square radius is defined as
\begin{equation}
\left\langle {r^2 } \right\rangle _{M,I = 0}  = \frac{3}{5}\frac{{\int\limits_0^\infty  {r^4 \sqrt { - g} B^0 4\pi r^2 dr} }}{{\left\langle {r^2 } \right\rangle _{I = 0} }}. 
\end{equation}
The other components of baryon's current are found 
\begin{equation}
B^i  = \frac{{i\varepsilon ^{ijk} }}{{2\pi ^2 }}\frac{1}{{\sqrt { - g} }}\frac{{\sin ^2 nf}}{{r^2 }}f'x_k Tr\left[ {\dot A^{ - 1} A\tau _j } \right].
\end{equation}
From [3], we can get
\begin{equation}
B^i  = \frac{{\varepsilon ^{ijk} }}{{2\pi ^2 }}\frac{1}{{\sqrt { - g} }}\frac{{\sin ^2 nf}}{{r^2 }}f'x_k \frac{{J_j }}{{2\Gamma }}.
\end{equation}
The time component of vector current $V$ is defined as
\begin{equation}
V_i^0  = \frac{i}{3}\frac{{\sin ^2 nf}}{{n^2 r^2 }}\left[ {F_\pi ^2  + \frac{4}{{e^2 }}\left( {Cf'^2  + \frac{{\sin ^2 nf}}{{n^2 r^2 }}} \right)} \right]Tr\left( {\dot AA^{ - 1} \tau _i } \right),
\end{equation}
or 
\begin{equation}
V_i^0  = \frac{{\sin ^2 nf}}{{3n^2 r^2 }}\left[ {F_\pi ^2  + \frac{4}{{e^2 }}\left( {Cf'^2  + \frac{{\sin ^2 nf}}{{n^2 r^2 }}} \right)} \right]\frac{{T_i }}{{2\Gamma }}.
\end{equation}
The density of charge of nucleon is defined by Gell-Mann and Nishijima 
\begin{equation}
Q = \int\limits_0^\infty  {\rho _{nu} dr}  = I_3  + \frac{1}{2}B,
\end{equation}
where
\begin{equation}
B = \int\limits_0^\infty  {4\pi r^2 \sqrt { - g} } B^0 dr =  - \int\limits_0^\infty  {\frac{{2f'\sin ^2 nf}}{\pi }} dr,
\end{equation}
\begin{equation}
I_3  = \int\limits_0^\infty  {\sqrt { - g} } r^2 V_3^0 d\Omega dr = \int\limits_0^\infty  {N\frac{{4\pi \sin ^2 nf}}{{3n^2 }}} \left[ {F_\pi ^2  + \frac{4}{{e^2 }}\left( {Cf'^2  + \frac{{\sin ^2 nf}}{{n^2 r^2 }}} \right)} \right]\frac{{T_3 }}{{2\Gamma }}dr.
\end{equation}
So
\begin{equation}
\rho _{nu}  = N\frac{{4\pi \sin ^2 nf}}{{3n^2 }}\left[ {F_\pi ^2  + \frac{4}{{e^2 }}\left( {Cf'^2  + \frac{{\sin ^2 nf}}{{n^2 r^2 }}} \right)} \right]\frac{{T_3 }}{{2\Gamma }} - \frac{{f'\sin ^2 nf}}{\pi }.
\end{equation}
For $T_3 = 1/2$ and $T_3 = -1/2$ we get respectively the density of charge of proton and neutron.  
The isoscalar magnetic moment and isovector magnetic moment are defined as
\begin{equation}
\vec \mu _{I = 0}  = \frac{1}{2}\int {\sqrt { - g} } \left( {\vec r \times \vec B} \right)4\pi r^2 dr,
\end{equation}
\begin{equation}
\mu _{I = 1}  = \frac{1}{2}\int {\sqrt { - g} \left( {\vec r \times \vec V^3 } \right)} 4\pi r{}^2dr. 
\end{equation}
According to [3], final results of Eqs. (113), (114) are defined
\begin{equation}
\mu _p  = \frac{1}{4}\left[ {\frac{4}{9}M_N \left( {M_\Delta   - M_N } \right)\left\langle {r^2 } \right\rangle _{I = 0}  + \frac{{2M_N }}{{M_\Delta   - M_N }}} \right],
\end{equation}
\begin{equation}
\mu _p  = \frac{1}{4}\left[ {\frac{4}{9}M_N \left( {M_\Delta   - M_N } \right)\left\langle {r^2 } \right\rangle _{I = 0}  - \frac{{2M_N }}{{M_\Delta   - M_N }}} \right].
\end{equation}

\section{Conclusions}
We have constructed the term of mass of pion and calculated numerically in a case of $n=4$. We have applied this model to define static properties of nucleon. A few obtained results are better than ones that predicted before [1, 3, 4, 9] as the axial coupling $g_A=$$\bf 1.22$ (in [1]: $g_A=1.3$, in [3]: $g_A=0.61$, in [4]: $g_A=0.65$, in [9]: $g_A=1.27$),  and  $\mu _{neu} (mag)=$$\bf -1.57$ (in [1]: $\mu _{neu} (mag)=-1.17$, in [3]: $\mu _{neu} (mag)=-1.31$, in [4]: $\mu _{neu} (mag)=-1.24$).
We have constructed successfully the formalism of supersymmetric composite Skyrme model. The proof was that the numerical solutions of field equation (24) were obtained. This was a role to calculate some static propertities of nucleon.
The formalism of gravitational composite Skyrme model was constructed. In fact, we have not calculated numerically the field equation and some properties of nucleon, but we believed that they will be perfomed in the near future. We also believed that our results could be applied in some fields of cosmology.

\newpage
\bf {* figures for a case of pion's mass}
\begin{figure}[ht]
\centering

\includegraphics[scale=0.55]{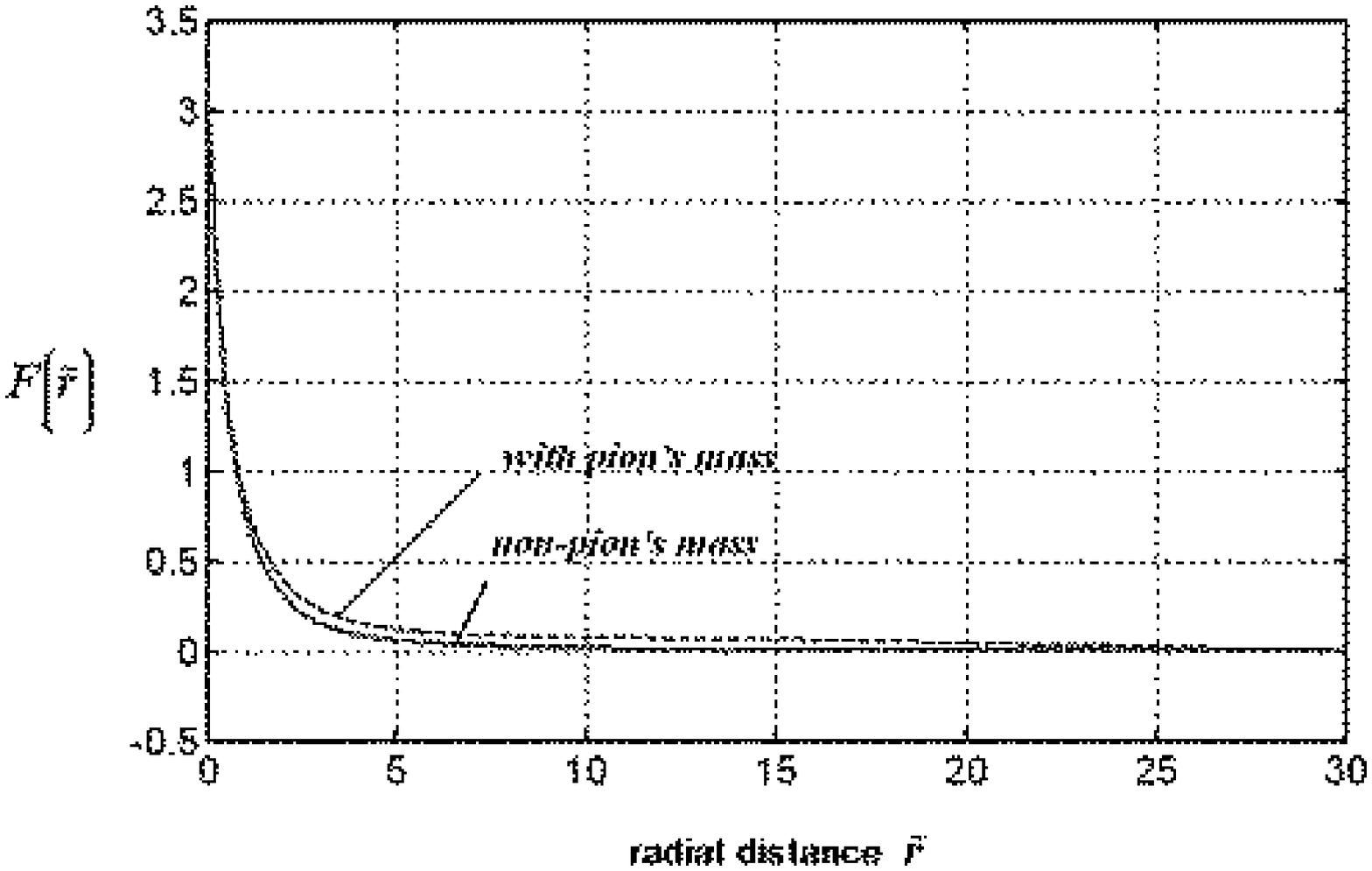}

\caption{{\it\small $F\left( \tilde r \right)$ as a function of the radial distance $\tilde r$}}

\end{figure}

\newpage

\begin{figure}[ht]

\centering

\includegraphics[scale=0.55]{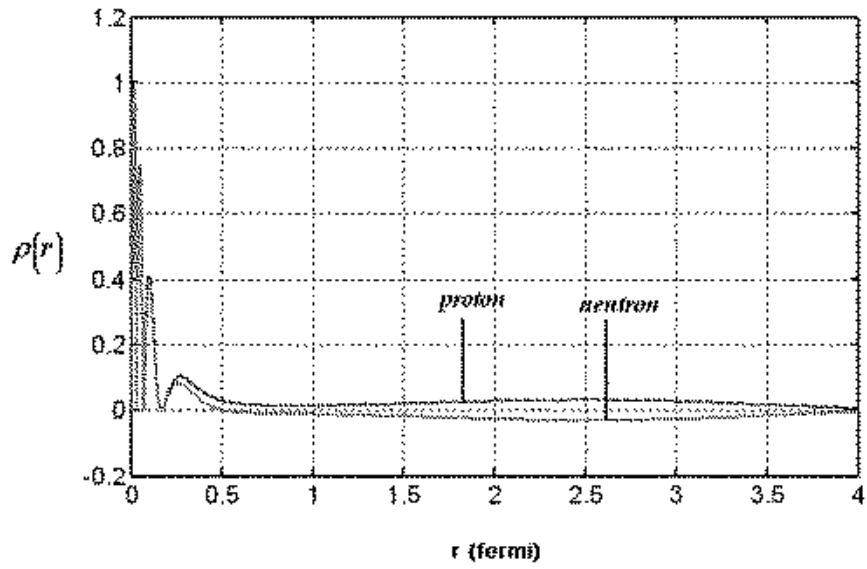}

\caption{{\it\small Neutron and proton charge densities $\rho
\left( r \right)$ as functions of the distance $\it r (fm)$}}

\end{figure}

\newpage
\bf{** the figure for a case of supersymmetry

\begin{figure}[ht]

\centering

\includegraphics[scale=0.55]{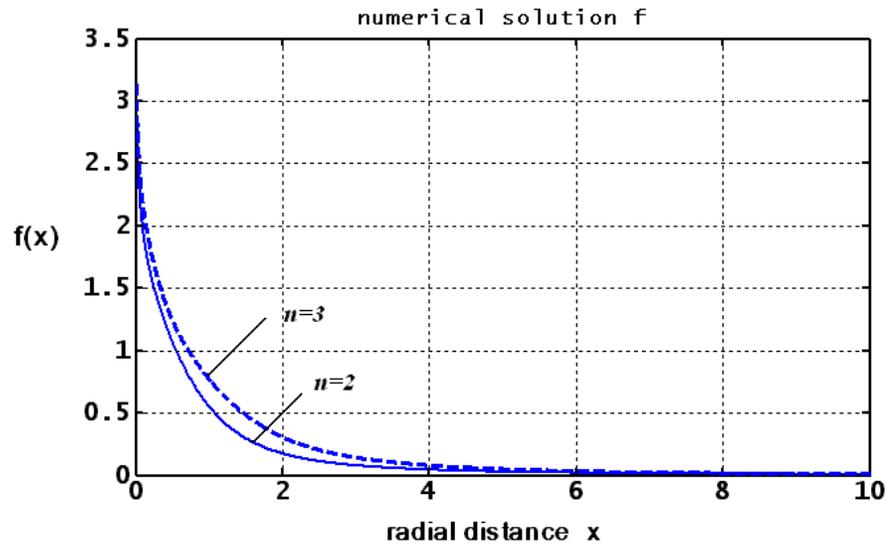}

\caption{{\it\small The numerical solution $f(x)$ at $\alpha  = 1,\beta  = 0, n = 2, 3$}}

\end{figure}

\end{document}